\documentclass{llncs}
\usepackage{amsmath,amssymb,latexsym,xspace,psfrag,graphicx}
\usepackage{fullpage}

\newcommand{\scO}{\mathcal{O}}
\newcommand{\scX}{\mathcal{X}}

\newcommand{\1}{\mathbf{1}}

\newcommand{\biggiven}{\;\big|\;}
\newcommand{\bigggiven}{\;\bigg|\;}

\newcommand{\Bi}{\operatorname{B}}

\renewcommand{\Pr}{\operatorname{Pr}}

\def\a{\alpha}

\def\d{\delta}

\def\e{\epsilon}

\newcommand{\proofend}{\hspace*{\fill}\mbox{$\Box$}}

\def\ER{{Erd\H{o}s-R\'{e}nyi}\xspace}

\title{Bias reduction in traceroute sampling:\\
towards a more accurate map of the Internet
}

\author{
Abraham D. Flaxman\inst{1}
\and 
Juan Vera\inst{2}
}

\institute{
Microsoft Research\\
Redmond, WA\\
\email{abie@microsoft.com}
\and
Georgia Institute of Technology\\
Atlanta, GA\\
\email{jvera@cc.gatech.edu}
}

\begin{document}
\maketitle

\begin{abstract}
Traceroute sampling is an important technique in exploring the
internet router graph and the autonomous system graph.  Although it is
one of the primary techniques used in calculating statistics about the
internet, it can introduce bias that corrupts these estimates.  This
paper reports on a theoretical and experimental investigation of a new
technique to reduce the bias of traceroute sampling when estimating
the degree distribution.  We develop a new estimator for the degree of
a node in a traceroute-sampled graph; validate the estimator
theoretically in \ER graphs and, through computer experiments,
for a wider range of graphs; and apply it to produce a new picture of
the degree distribution of the autonomous system graph.
\end{abstract}

\section{Introduction}
The internet is quite a mysterious network.  It is a huge and complex
tangle of routers, wired together by millions of edges.  To understand
this \emph{router graph} is quite a challenge, one that has driven research
for the last decade.

%A clear picture of the this network can also guide the
%development of new network algorithms.

The router graph has a natural clustering into Autonomous Systems
(ASes), which are sets of routers under the same management.
Producing an accurate picture of the \emph{AS graph} is an important step
towards understanding the internet.

There are three techniques for finding large sets of edges in the AS
graph: the {\tt WHOIS} database, BGP tables, and traceroute sampling.
No approach is clearly superior, and the results of the different
approaches are compared in detail in a recent paper
\cite{MahadevanKFDcV2006}.

The present paper focuses on traceroute sampling, an approach applicable to the router graph as well as the AS graph.  Traceroute
sampling consists of recording the paths that packets follow when they
are sent from monitor nodes to target nodes, and merging all of these
paths to produce an approximation of the AS graph.

A seminal analysis using both traceroute sampling and BGP tables concluded
that the AS graph degree distribution follows a power-law (meaning
that the number of ASes of degree $k$ is proportional to $k^{-\a}$ for
a wide range of $k$ values) \cite{FaloutsosFF1999}.  This caused a
shift in simulation methodology for evaluating network algorithms and
also contributed to the avalanche of recently developed network models
which produce power-law degree distributions.
%(It was not without controversy; see for
%example, \cite{ChenCGJSW2002}.) 

However, the true nature of the AS-graph degree distribution was called
into question by computer experiments on synthetic graphs
\cite{LakhinaBCX2003,PetermannR2004}.  These experiments show that if
the sets of monitor and target nodes are too small then traceroute
sampling will produce a power-law degree distribution, even when the
underlying graph has a tightly concentrated degree distribution.
Theoretical follow-up work proved rigorously that in many
non-power-law graphs, including random regular graphs, an
idealized model of traceroute sampling yields power-law degree
distributions \cite{ClausetM2005,AchlioptasCKM2005}.

Subsequent computer experiments have led some to believe that the bias
inherent to traceroute sampling can be ignored, at least for making a
qualitative distinction between scale-free and homogeneous graphs, when
using a large enough set of monitor nodes \cite{GuillaumeLM2006}.
This is also supported by an analysis using the statistical physics
technique of mean field approximation \cite{DallastaABVV2005}.
%However, the mean field approximation may be inaccurate in graphs with
%local clustering, and the proposed benefits of a massively distributed
%monitor set seem to conflict with a previous study showing that
%adding more monitor nodes does not yield that much additional
%information \cite{BarfordBBC2001}.

\subsection{Our contribution}
This paper proposes a new way forward in the struggle to characterize
the degree distribution of the AS graph.  Our contribution has three parts:
\begin{enumerate}
\item We derive a statistical technique for reducing the bias in
traceroute sampling;
\item We verify the technique experimentally and theoretically, in the
  framework previously studied in~\cite{LakhinaBCX2003,ClausetM2005};
\item We use the traceroute bias-reduction technique to generate a
  more accurate picture of the AS degree distribution over time, which
  suggests that aspects of commercially available technology are
  reflected in the network topology.
\end{enumerate}
Our approach for reducing the bias in traceroute sampling is based on
a technique from biostatistics, the multiple-recapture census, which
has been developed for
estimating the size of an animal population \cite{PickandsR1987}
(this technique also has applications to proofreading \cite{Polya1975}).  
%It was pioneered for calculating waterfowl abundance on the basis of
%banding returns \cite{Lincoln1930}).
However, we do not have the benefit of independent random variables
which are central to the animal counting and proofreading statistics,
and so we must adapt the technique to apply to random variables with
complicated dependencies.

To provide some evidence that this bias-reduction technique actually
reduces bias, we consider a widely used model of traceroute sampling,
which assumes that data travels from monitor to target along the
shortest path in the network.  It is generally believed that the path
that data actually takes is \emph{not} the shortest path, but that the
shortest path is an acceptable approximation of the actual path (see
\cite{LeguayLFS2005} for a discussion of this approximation).  In this
model, it is possible to check theoretically and experimentally that
the bias reduction provides a better estimate of the degree
distribution.  We show that the new estimation is asymptotically unbiased for the
\ER random graph $G_{n,p}$ when $np \gg \log n$, and that it gives
improved estimates for finite instances from a variety of different graphs.

\begin{figure}
\begin{center}
\psfrag{a}[cl][cl]{\tiny 2004, Biased}
\psfrag{b}[cl][cl]{\tiny 2004, Bias Reduced}
\psfrag{c}[cl][cl]{\tiny 2002, Bias Reduced}
\psfrag{x}[cl][cl]{$k$}
\psfrag{y}[cl][cl]{$\Pr[\deg(u) > k]$}
\psfrag{0}[cl][cl]{\tiny $.1^{0}$}
\psfrag{1}[cl][cl]{\tiny $.1^{1}$}
\psfrag{2}[cl][cl]{\tiny $.1^{2}$}
\psfrag{3}[cl][cl]{\tiny $.1^{3}$}
\psfrag{4}[cl][cl]{\tiny $.1^{4}$}
\psfrag{p}[cl][cl]{\tiny $.1^{1}$}
\psfrag{q}[cl][cl]{\tiny $.1^{2}$}
\psfrag{u}[cl][cl]{\tiny $64$}
\psfrag{v}[cl][cl]{\tiny $90$}
\psfrag{A}[cl][cl]{\tiny $10^{0}$}
\psfrag{B}[cl][cl]{\tiny $10^{1}$}
\psfrag{C}[cl][cl]{\tiny $10^{2}$}
\psfrag{D}[cl][cl]{\tiny $10^{3}$}
\includegraphics[width=6.5in]{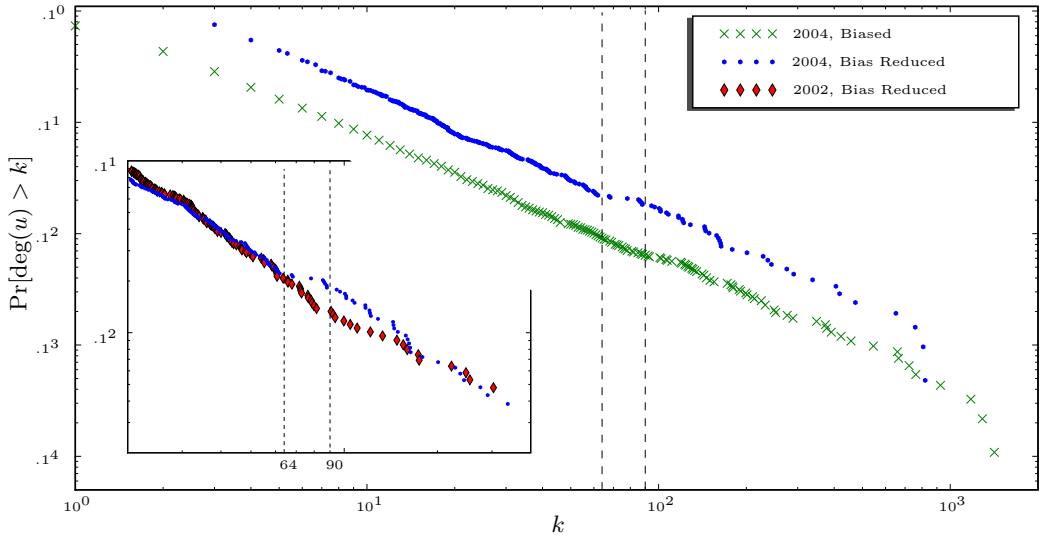}
\end{center}
\caption{Degree sequence ccdf estimates for the AS graph (from CAIDA
  skitter).  Main panel: March, 2004, with and without bias reduction.
  Inset: a portion of ccdf for March, 2004 and March, 2002, both with
  bias reduction.  The nodes with degree between 65 and 90 in 2002
  have disappeared in 2004.}
\label{fig:AS-degree-sequence}
\end{figure}

Finally, we use the bias-reduction technique on real data, traceroute
samples from the internet.  The new estimate of the AS-graph degree
distribution is still scale-free over two orders of magnitude, with an
exponent very similar to the uncorrected degree distribution (see
Figure \ref{fig:AS-degree-sequence}).  A by-product of bias reduction
is the removal of all vertices with degree less than 3, and this
increases the average degree.  For example, in March 2004 (the month
used for comparison in \cite{MahadevanKFDcV2006}), the biased estimate
of average degree is $6.29$, while after bias reduction the average
degree is $12.66$ (which is very close to $12.52$, the biased average
degree when restricted to vertices of degree at least $3$).  An interesting
feature in the bias-reduced AS degree distribution (from March 2004)
is the lack of nodes with degree between 65 and 90; at the time, a
popular router maker offered a router which provided up to $64$ ports
per chassis.  In March 2002, before this product was available,
% Juniper T320 edge router was available, 
there was no dearth of 65 degree nodes.

\subsection{Related work}
Internet mapping by traceroute sampling was pioneered by Pansiot and
Grad in \cite{PansiotG1998}, and the scale-free nature of the degree
distribution was observed by Faloutsos, Faloutsos, and Faloutsos in
\cite{FaloutsosFF1999}.  Since 1998, the Cooperative Association for
Internet Data Analysis (CAIDA) project \emph{skitter} has
archived traceroute data that is collected daily \cite{claffyMM1999}.  The bias
introduced by traceroute sampling was identified in computer
experiments by Lakhina, Byers, Crovella, and Xie in
\cite{LakhinaBCX2003} and Petermann and De Los Rios
\cite{PetermannR2004}, and formally proven to hold in a model of one-monitor,
all-target traceroute sample by Clauset and Moore
\cite{ClausetM2005} and, in further generality, by Achlioptas,
Clauset, Kempe, and Moore \cite{AchlioptasCKM2005}.  Computer
experiments by to Guillaume, Latapy, and Magoni \cite{GuillaumeLM2006}
and an analysis using the mean field approximation of statistical
physics due to Dall'Asta, Alvarez-Hamelin, Barrat, V\'{a}zquez, and
Vespignani \cite{DallastaABVV2005} argue that, despite the bias
introduced by traceroute sampling, some sort of scale-free behavior
can be inferred from the union of traceroute-sampled paths.

The present paper provides a new avenue for investigating these
controversial questions, by developing a method for \emph{correcting}
the bias introduced by traceroute sampling.  Another recent paper by
Viger, Barrat, Dall'Asta, Zhang and Kolaczyk applied techniques from
statistics to reduce the bias of traceroute sampling
\cite{VigerBDZK2005}.  That paper focused on estimating the number of
nodes in the AS graph, and applied techniques from a different problem
in biostatistics, estimating the number of species in a bioregion.
The problem of correcting bias in sampled networks has a long history
in sociology, although the biases in that domain seem somewhat
different; see the surveys by Frank, by Klovdahl, or by Salganik and
Heckathorn for an overview
\cite{Frank1981,Klovdahl1989,SalganikH2004}.
%Similar in spirit, but not in practice, to
%\cite{StutzbachRDSW2006,GarlaschelliL2007}. 

In addition to traceroute sampling, maps of the AS graph have been
generated in two different ways, using BGP tables and using the {\tt WHOIS}
database.  A recent paper by Mahadevan, Krioukov, Fomenkov,
Dimitropoulos, claffy, and Vahdat provides a detailed comparison of
the graphs that result from each of these measurement techniques
\cite{MahadevanKFDcV2006}.

\subsection{Outline of what follows}
The new estimator for the degree of a node in the AS graph is
developed from multiple-recapture population estimation in Section
\ref{sec:estimation technique}.  Section \ref{sec:proof} argues that
this estimator generates an asymptotically unbiased degree
distribution for the \ER graph $G_{n,p}$ when $p \gg \log n$, which
rigorously demonstrates that the new estimator can reject a null
hypothesis.
% and give formal confidence that the AS graph is not a
%uniformly random graph with given number of nodes and edges.
Section \ref{sec:computer experiments} presents additional evidence
that the new estimator reduces the bias of traceroute sampling, in the
form of computer experiments on synthetic networks.  Section
\ref{sec:as graph} provides a comparison between the degree sequence
predicted by the new estimator and the previous technique, and details
how, after bias reduction, the degree distribution may reflect
economic and technological factors present in the system,
i.e., there a significantly larger marginal cost of adding a 65th
neighbor than adding a 64th neighbor when using the Juniper T320 edge
router.  Section \ref{sec:end} provides a conclusion and focuses on
directions of future research to strengthen this approach.

\section{Estimation Technique}
\label{sec:estimation technique}
The classical capture-recapture approach to estimating an animal
population has two phases.  First, an experimenter captures animals
for a given time period, marks them (with tags or bands), and
releases them, recording the total number of animals captured.  Then,
the experimenter captures animals for a second time period, and
records both the number of animals recaptured and the total number of
animals captured during the second period.  If $A$ denotes the number
of animals captured in phase one, $B$ denotes the number captured during
phase two, and $C$ denotes the number captured in phase one and
captured again in phase two, then an
estimate of total population size is given by
\[
\widehat{N} = 
\begin{cases}
\frac{A B}{C}, &\quad \text{if } C \neq 0;\\
\infty, &\quad \text{otherwise}.
\end{cases}
\]

If the true population size is $N$, and each animal is captured or not
captured during each phase independently, with probability $p_1$ during
phase one and probability $p_2$ during phase two, then $\widehat{N}$ is the
maximum likelihood estimate of $N$ \cite{PickandsR1987}.  For more
than two phases, the maximum likelihood estimator does not have a simple
closed form, but it can be computed efficiently using the techniques developed in
\cite{PickandsR1987}.

When estimating the degree of a particular AS by traceroute sampling,
each edge corresponds to an animal, and each monitor node corresponds
to a recapture phase.  Unfortunately, in this setting there is no
reason to believe that the events ``monitor $i$ observes edge $j$'' are
independent.  Indeed, when shortest-path routing is used (as an
approximation of BGP routing), these events are highly dependent.
However, it is still possible adapt the capture-recapture estimate to
reduce bias in this case.

Let $G$ be a graph, and let $s$ and $t$ be monitor nodes in $G$.  Let
$G_s$ be the union of all routes discovered when sending packets from
$s$ to every node in the target set.  Define $G_t$ analogously.  Let
$N_s(u)$ denote the neighbors of $u$ in $G_s$ and define $N_t(u)$
analogously.

Using this notation, the  modification of the capture-recapture
estimate proposed for traceroute sampling is given by
\[
\widehat{\deg}_{s,t}(u) = \begin{cases}
\frac{|N_s(u)|\cdot|N_t(u)|}{|N_s(u) \cap N_t(u)|},
&\quad 
\text{if } |N_s(u) \cap N_t(u)| > 2;
\\
\infty, &\quad \text{otherwise.}
\end{cases}
\]

When more than 2 monitor nodes are available, pair up the monitors,
consider the estimates given by each pair that are not $\infty$, and
for the final estimator, use the median of these values.  So, if the
monitor nodes are paired up as $(s_1,t_1), (s_2,t_2), ..., (s_k,t_k)$
then
\[
\widehat{\deg}(u) = 
\operatorname{median}
\left(\left\{
\widehat{\deg}_{s_i,t_i}(u)
\neq \infty
\right\}\right).
\]

This degree estimator can also provide an estimate of the cdf of the
degree distribution (i.e., the fraction of nodes with degree at most
$k$) according to the formula
\[
\widehat{d_{\leq k}} = \widehat{\Pr}[\deg(u) \leq k] = \frac
{\# \{u :  \widehat{\deg}_{s,t}(u) \leq k \}}
{\# \{u : \widehat{\deg}_{s,t}(u) < \infty \}}.
 \]

\emph{Discussion:} It may seem wasteful to consider the median-of-two-monitors estimate instead of combining all available monitors in a
more holistic manner.  However, we have conducted computer experiments
with maximum likelihood estimators for multiple-recapture population
estimation with more than two phases, and the adaptations we have
considered thus far perform significantly worse than the
median-of-two-monitors approach above.  This is probably due to the
complicated dependencies of several overlapping shortest-path trees.
However, the exploration we have conducted to date is not exhaustive,
and does not rule out the possibility that a significantly better
estimator exists.

\section{Theoretical analysis}
\label{sec:proof}
This section and the next intend to provide some assurance that repeated
application of $\widehat{\deg}(u)$ is an accurate way to estimate
the degree distribution of the sampled graph.

This section provides a theoretical analysis of the performance of
$\widehat{\deg}(u)$ in a very specific setting: when the underlying graph is the \ER graph
$G_{n,p}$ with $n$ sufficiently large, $np \gg \log n$, and every
vertex is a target node.  For the
purpose of analysis, this section and the next assume that traceroute
finds a shortest path from monitor to target.  This is the same setting
that is considered in \cite{ClausetM2005}.

\begin{theorem}
Let $G \sim G_{n,p}$ be a random graph with $np = d \gg \log n$,
and let $s,t,$ and $u$ be uniformly random vertices of $G$.  Then, for any $k$,
with high probability,
\[
\widehat{d_{\leq k}} = \frac
{\# \{u :  \widehat{\deg}(u) \leq k \}}
{\# \{u : \widehat{\deg}(u) < \infty \}}
 =
\frac
{\# \{u : \deg(u) \leq k\}}
{n}
 \pm \scO\left(1/d\right).
\]
\end{theorem}

\emph{Proof sketch:}
The analysis \emph{two} breadth-first-search trees in a random graph
is difficult when the average degree is small.  But, for $d$
moderately large, as in this theorem, the situation is simpler.

It follows from the branching-process approximation of breadth-first
search that with high probability there are $(1\pm\e)d^i$ vertices at distance exactly
$i$ from $s$ (or $t$) when $i < (\log n)/(\log d)$. Thus, almost
all vertices are distance $\lceil(\log n)/(\log d)\rceil$ apart.  For
ease of analysis, suppose that $\ell = (\log n)/(\log d)$ is an
integer.

So, with high probability, if $u$ is at distance $\ell$ from $s$ or $t$ then it is a leaf
node in $G_s$ or $G_t$.  In this case, $|N_s(u) \cap N_t(u)| \leq 1$
and therefore $\widehat{\deg}(u) = \infty$.

Now, consider the case where vertex $u$ is distance $i$ from $s$ and
distance $j$ from $t$, where $i,j < \ell$.  Let $N(u)$ denote the
neighbors of $u$ in $G$, and then let $S$ be the set of vertices
within distance $i$ of $s$ in $G$ and let $T$ be the set of vertices
within distance $j$ of $t$ in $G$.  Conditioned on $S$, $T$ and
$N(u)$, the set of indicator random variables 
\[\bigg\{ 
\1[v\in N_s(u)],
\1[v\in N_t(u)]
\colon v \in N(u)\setminus (S \cup T)
\bigg\}
\]
is independent, and, for $v \in N(u)\setminus (S\cup T)$, $\Pr[v\in
N_s(u)]$ and $\Pr[v\in N_t(u)]$ are functions of $S$ and $T$, but
constants with respect to $v$, i.e., $\Pr[v\in N_s] = p_s$ and
$\Pr[v\in N_t] = p_t$.  So, besides any edges between $u$ and
$S\cup T$, the edges
incident to $u$ in $G_s[S]$ and $G_t[T]$ yield the random variables
$|N_s(u)|$, $|N_t(u)|$, and $|N_s(u) \cap N_t(u)|$, which correspond to
$A$, $B$, and $C$ in the capture-recapture estimate of population
size.  For example, if there is only one edge incident to $u$ in
$G_s[S]$ and only one in $G_t[T]$, and these edges are different, then
\[
\Pr\left[\widehat{\deg(u)} \geq k \bigggiven S, T, N(u)\right]
= 
\Pr\left[\frac{(A+1)(B+1)}{C} \geq k\right],
\]
where $C \sim \Bi(|N(u)|-2,p_s p_t)$, $A \sim C + \Bi(|N(u)|-1-C,p_s)$, and $B
\sim C + \Bi(|N(u)|-1-C,p_t)$.  If $k$ is sufficiently large and $p_s$ and
$p_t$
are not too small then this probability is concentrated in the range $k
= |N(u)| \pm \sqrt{|N(u)|}$.

To complete the proof, it remains to show that, with probability
$1-\scO(1/d)$, $p_s, p_t \geq \e$ and $|N(u) \cap (S \cup T)| \leq 2$,
and from this show that, for $A,B,C$ defined analogously to above,
\[
\Pr\left[\frac{(A+1)(B+1)}{C} \geq k\right]
= \Pr[|N(u)| \geq k] + \scO(1/d).
\]
\proofend

\emph{Discussion:}  
This analysis would go through without modification if the estimate also included
samples where $|N_s(u) \cap N_t(u)| = 2$, but the definition of
$\widehat{\deg}(u)$ from above seems to behave better under finite
scaling.

The proof sketch can be adapted for random
graphs with other degree distributions, provided that the average
degree is large.  However, the proof relies on the fact that the graph is
\emph{locally tree-like}, which ensures that $N(u) \cap (S \cup T)$ is
likely to be small.  This assumption does not seem to hold in the AS
graph, and even $G_s$, the union of all routes discovered from a single
monitor node $s$, has some triangles.  The next section includes
evidence from computer experiments that in graphs which are \emph{not}
locally tree-like, such as the random geometric graph, estimator
$\widehat{\deg}(u)$ is not asymptotically unbiased, but can still
reduce some amount of bias.  Proving this rigorously may
be a difficult task.

\section{Computer experiments}
\label{sec:computer experiments}

This section describes the results of a series of computer experiments
conducted to investigate how well $\widehat{d_{\leq k}}$ approximates the
true degree distribution.

We consider three different distributions for random graphs, the \ER
model, the Preferential Attachment model, and the random geometric
graph.  Additionally, we consider synthetic data based on a real-world
graph, the Western States Power Grid (WSPG), which Duncan Watts has graciously
made available to researchers \cite{WattsS1998}.  These graphs will all
be described in more detail below.

For each graph, we set edge $e$ to be of length $1+\eta_e$, where
$\eta_e$ is selected uniformly from the interval $[-1/n,1/n]$, where
$n$ is the number of vertices.  This 
ensures that there are not multiple shortest paths between pairs of
vertices.  We approximate the path that data takes from a monitor to
a target node by the shortest path.  This follows the experimental design of
\cite{LakhinaBCX2003}.

For each graph distribution, and for a range of graph sizes, edge
densities, monitor set sizes, and target set sizes, we estimate the
degree of every vertex by $\widehat{\deg}(u)$ and by the biased
estimator given by the union of the edges discovered by traceroute
sampling,
\[
\widehat{\deg}_{\text{biased}}(u) = \left| \bigcup_{s \in V_m} N_s(u)\right|,
\]
where $V_m$ is the set of monitor nodes and $N_s(u)$ denotes the
neighbors of $u$ in the union of all routes discovered when sending
packets from $s$ to every node in the target set $V_t$.  This
provides estimates of the degree distribution cdf, by the reduced bias
estimator 
$\widehat{d_{\leq k}}$ from above and by the biased
estimator $\widehat{d_{\leq k}}_{\text{biased}}$, defined by
\[
\widehat{d_{\leq k}}_{\text{biased}}
= \frac
{\# \{u :  \widehat{\deg}_{\text{biased}}(u) \leq k \}}
{\# \{u : \widehat{\deg}_{\text{biased}}(u) \geq 1 \}}.
\]
$\widehat{d_{\leq k}}_{\text{biased}}$ has been the primary approach considered in
prior work.

We use these
estimates to calculate the
$\ell_2$ error of the degree distribution cdf estimate, given by
\[
\text{err}_{\text{biased}} = 
\frac
{\left(\sum_{k = 0}^{\infty} 
\left( \widehat{d_{\leq k}}_{\text{biased}}
 - \Pr[\deg(u) \leq k] \right)^2\right)^{1/2}}
{\left(\sum_{k = 0}^{\infty} 
\Pr[\deg(u) \leq k]^2\right)^{1/2}}
,
\]
and
\[
\text{err}_{\text{reduced}} = 
\frac
{\left(\sum_{k = 0}^{\infty} 
\left( \widehat{d_{\leq k}}
 - \Pr[\deg(u) \leq k] \right)^2\right)^{1/2}}
{\left(\sum_{k = 0}^{\infty} 
\Pr[\deg(u) \leq k]^2\right)^{1/2}}
,
\]
where $\Pr[\deg(u) \leq k] = {\# \{u : \deg(u) \leq k \}}/n$ is the
probability with respect to a uniformly random choice of $u$ from the
vertices of $G$.

We also exhibit plots of the distribution and the two estimates
for a typical parameter setting.  All error values reported are the
median value of $100$ experiments, and the plots show the distribution
with the median error as well as the pointwise $90$th percentile
values from the $100$ experiments.

\subsection{Random graph, $G_{n,m}$}
\begin{figure}
\begin{tabular}{cc}
\psfrag{True Distribution}[cl][cl]{\tiny True Dist.}
\psfrag{Biased Estimate}[cl][cl]{\tiny Biased Est.}
\psfrag{Bias Reduced}[cl][cl]{\tiny Bias Reduced}
\includegraphics[width=3.0in]{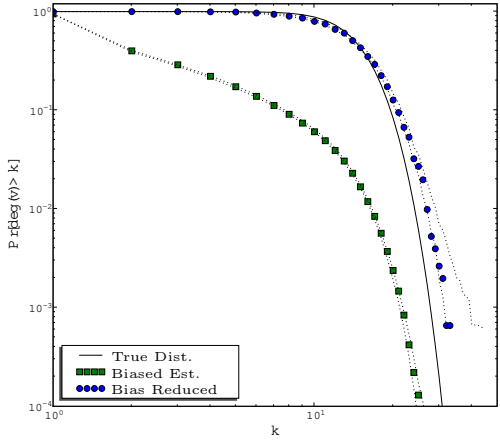}
&
\psfrag{True Distribution}[cl][cl]{\tiny True Dist.}
\psfrag{Biased Estimate}[cl][cl]{\tiny Biased Est.}
\psfrag{Bias Reduced}[cl][cl]{\tiny Bias Reduced}
\includegraphics[width=3.0in]{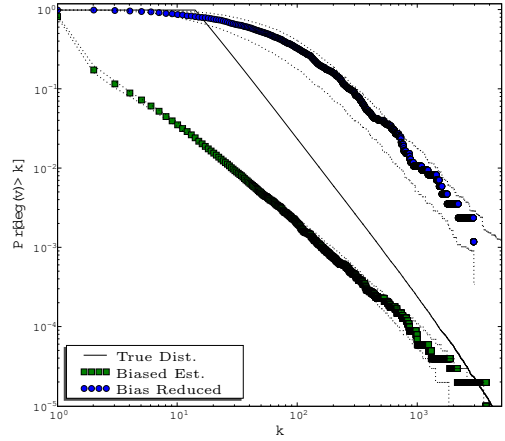}\\
(a) $G_{n,m}$ with $n = 100,000$, $d = 2m/n = 15$.
&
(b) PA graph with $n = 100,000$, $m = 15$.\\
\psfrag{True Distribution}[cl][cl]{\tiny True Dist.}
\psfrag{Biased Estimate}[cl][cl]{\tiny Biased Est.}
\psfrag{Bias Reduced}[cl][cl]{\tiny Bias Reduced}
\includegraphics[width=3.0in]{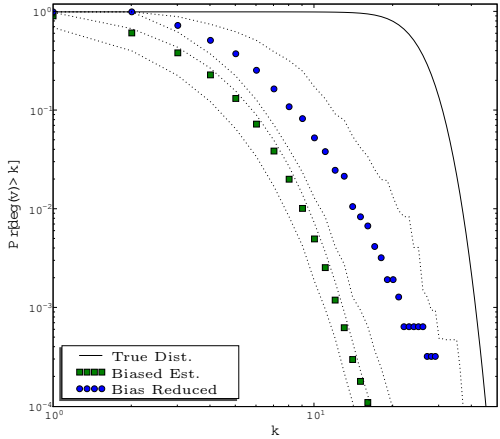}
&
\psfrag{True Distribution}[cl][cl]{\tiny True Dist.}
\psfrag{Biased Estimate}[cl][cl]{\tiny Biased Est.}
\psfrag{Bias Reduced}[cl][cl]{\tiny Bias Reduced}
\includegraphics[width=3.0in]{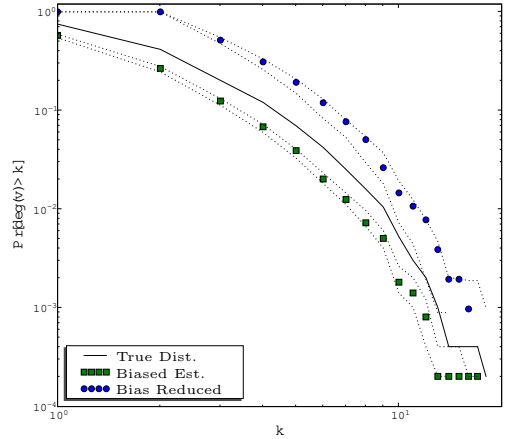}\\
(c) $G(\scX; r)$ with $n = 100,000$, $d =
\pi r^2 = 25$.
&
(d) Western states power graph from \cite{WattsS1998}.
\end{tabular}
\caption{Degree sequence ccdf, biased, and bias reduced
  estimators for synthetic data, with $2$ monitor nodes chosen
  uniformly at random, $n$ target nodes, and shortest path sampling
  used to approximate traceroute.  Plots based on 100 trials, where
  data points correspond to trial with median $\ell_2$ error, and
  dotted region shows pointwise bounds on $90\%$ of trials.}
\label{fig:sim}
\end{figure}

The \ER distribution of graphs, $G_{n,m}$, can be generated by
choosing a graph uniformly at random from all graphs with $n$ vertices
and $m$ edges \cite{ErdosR1959}.  It was not developed to model
real-world graphs, but it is analytically tractable and can provide
insight into the behavior of more realistic graph models.  It can also
be used as a null hypothesis.  Section \ref{sec:proof} proved that
$\widehat{\deg}(u)$ and $\widehat{d_{\leq k}}$ are asymptotically
unbiased for $G_{n,p}$ when $np \gg \log n$.  Conventional
wisdom holds that anything true for $G_{n,p}$ is also true for
$G_{n,m}$ when $m \approx \binom{n}{2}p$, and computer experiments
support this conclusion, even for moderately size $n$ and $m$, as
shown in Table \ref{table:Gnp} and Figure \ref{fig:sim}a.  These
experiments indicate that $\widehat{\deg}(u)$ and $\widehat{d_{\leq
k}}$ are also good estimators when the number of targets $n_t$ is a
reasonably small fraction of $n$, which is the case in traceroute
sampling of the AS graph.

\begin{table}
\begin{center}
\begin{tabular}{|r|r|r|r|r|r|}
\hline
$n$     &$d$
            &$n_m$&$n_t$ & \% $\text{err}_{\text{biased}}$ 
                                & \% $\text{err}_{\text{reduced}}$\\
\hline
  1,000 & 15&  2 & $n/8$ & 3.38 & 3.15 \\
        &   &    & $n/2$ & 3.08 & 0.96 \\
        &   &    & $n$   & 2.81 & 0.42 \\
        &   &  8 & $n/2$ & 2.11 & 0.81 \\
        &   & 16 & $n/2$ & 1.38 & 0.80 \\
\hline
 10,000 & 20&  2 & $n/8$ & 4.02 & 2.10 \\
        &   &    & $n/2$ & 3.75 & 1.25 \\
        &   &    & $n$   & 3.51 & 0.46 \\
\hline
100,000 & 15&  2 & $n$   & 2.81 & 0.21 \\
\hline
\end{tabular}
\end{center}
\caption{$\ell_2$ error in degree distribution
  estimation with and without bias reduction for \ER graph, $G_{n,m}$
  where $d = 2m/n$, with $n_m$ monitors and $n_t$ targets (median values
  of $100$ trials).}
\label{table:Gnp}
\end{table}

\subsection{Preferential Attachment Graph}
The preferential attachment (PA) graph was proposed for a model of the
internet and the world wide web by Barab\'{a}si and Albert in
\cite{BarabasiA1999}, and this has generated a large body of
subsequent research, although the validity of the model as a
representation of the router graph or the AS graph has been questioned
(see, for example, \cite{ChenCGJSW2002}).  The estimator
$\widehat{\d_{\leq k}}$ does not perform particularly well on the PA
graphs that we used in our experiments, generating $\ell_2$ error that
is sometimes smaller and sometimes larger than the biased estimator
(see Table \ref{table:pa}).

The most interesting detail of this series of
experiments is the shape of the degree distribution estimated by
$\widehat{\d_{\leq k}}$.  When plotted on a log-log scale (Figure
\ref{fig:sim}b), the biased estimate of the degree distribution appears
to be straight line, although with a different slope than the
underlying distribution (this is consistent with the theoretical
results of \cite{AchlioptasCKM2005}).  However, the
``biased reduced'' estimate appears to fall off faster than linear
(when plotted on a log-log scale).
This is typical of the experiments we conducted with other
parameter settings for the PA graph.  It could
be an effect of the instance sizes being too small, but it persists over
two orders of magnitude.  Thus, it seems that 
locally non-tree-like aspects of the PA graph are decreasing the accuracy
of  $\widehat{\d_{\leq k}}$.  As shown in Figure
\ref{fig:AS-degree-sequence} and to be elaborated upon in Section
\ref{sec:as graph}, the degree distribution of the AS graph \emph{does
  not} fall off faster than linear when estimated with
$\widehat{\d_{\leq k}}$.  This could mean that the shortest path
routing used in the experiment is not a close enough approximation of
the true traceroute sampled paths.  But it \emph{could} be interpreted
as additional evidence that the AS graph is not distributed according
to the PA graph process.

\begin{table}[h]
\begin{center}
\begin{tabular}{|r|r|r|r|r|r|}
\hline
$n$     &$m$
            &$n_m$&$n_t$ & \% $\text{err}_{\text{biased}}$ 
                                & \% $\text{err}_{\text{reduced}}$\\
\hline
  1,000 & 5 &  2 & $n/8$ & 2.29 & 2.35 \\
        &   &    & $n/2$ & 1.95 & 2.66 \\
        &   &    & $n$   & 1.71 & 2.88 \\
        &   &  8 & $n/2$ & 1.26 & 2.00 \\
        &   & 16 & $n/2$ & 0.91 & 1.57 \\
\hline
 10,000 & 10&  2 & $n/8$ & 3.47 & 2.36 \\
        &   &    & $n/2$ & 3.23 & 3.39 \\
        &   &    & $n$   & 3.03 & 4.31 \\
\hline
100,000 & 15&  2 & $n$   & 3.99 & 4.43 \\
\hline
\end{tabular}
\end{center}
\caption{$\ell_2$ error in degree distribution estimation with and
  without bias reduction for Preferential Attachment graph with $n$
  nodes and $m$ out-edges per node, $n_m$ monitors and $n_t$
  targets (median values of $100$ trials).}
\label{table:pa}
\end{table}

\subsection{Random Geometric Graph, $G(\scX;r)$}
For graphs with high clustering coefficient, the proof sketched in
Section \ref{sec:proof} will not apply.  However, the traceroute paths
found by skitter exhibit some level of clustering.  To investigate
the performance of the bias-reduction technique on graphs with
clustering, we examine random geometric graphs $G(\scX;r)$.  These
graphs are formed by selecting a set of $n$ points independently and
uniformly at random from the unit square, and linking two points with
an edge if and only if they are within $\ell_2$ distance $r$ (for a
detailed treatment, see \cite{Penrose2003}).  The performance of the
bias-reduction technique is summarized for a variety of geometric
random graphs in Table \ref{table:geo}.

The plot exhibited in Figure \ref{fig:sim}c is typical for the
performance of bias reduction on random geometric graphs; although the
bias-reduced estimate is closer to the truth, it is still quite far
away from it.  The tail of the estimated ccdf, with or without bias
reduction, falls off noticeably more slowly than that of the true degree
distribution, and looks more like a power-law than it should.

In light of this, it seems that future research should investigate the
amount of clustering present in the AS graph.  This will permit us to
better gauge the accuracy of the bias-reduced estimate of the degree
distribution there.  However, understanding clustering in the AS
graph is hard for the same reasons that understanding the degree
distribution is hard, which is due to the lack of unbiased data.

\begin{table}[ht]
\begin{center}
\begin{tabular}{|r|r|r|r|r|r|}
\hline
$n$     &$d$
            &$n_m$&$n_t$ & $\text{err}_{\text{biased}}$ 
                                & $\text{err}_{\text{reduced}}$\\
\hline
  1,000 & 15&  2 & $n/8$ & 3.14 & 2.77 \\
        &   &    & $n/2$ & 2.91 & 2.49 \\
        &   &    & $n$   & 2.73 & 2.17 \\
        &   &  8 & $n/2$ & 2.45 & 2.50 \\
        &   & 16 & $n/2$ & 2.23 & 2.49 \\
\hline
 10,000 & 20&  2 & $n/8$ & 3.87 & 3.57 \\
        &   &    & $n/2$ & 3.68 & 3.36 \\
        &   &    & $n$   & 3.55 & 3.16 \\
\hline
100,000 & 25&  2 & $n$   & 4.19 & 3.90 \\
\hline
\end{tabular}
\end{center}
\caption{$\ell_2$ error in degree distribution
  estimation with and without bias reduction for geometric random
  graph, $G(\scX, r)$
  where $d = \pi r^2 n$, with $n_m$ monitors and $n_t$ targets (median values
  of $100$ trials).}
\label{table:geo}
\end{table}

\subsection{Western States Power Graph}

In addition to studying the behavior of bias reduction
on the random graphs describe above, we also consider the
estimator's performance on synthetic data that is based on a network
from the real world, the Western States Power Graph (WSP
Graph) \cite{WattsS1998}.  This graph represents the power
transmission links between $4,941$ nodes, representing the generators,
transformers, and substations in the Western United States.  It is roughly
similar in size to the AS graph, and also similar because both networks
represent real objects which are connected by real wires.

The result of the bias-reduction technique is shown in Figure
\ref{fig:sim}d.  The $\ell_2$ error is higher after bias reduction,
but this is because the bias-reduction technique filters out all
vertices of degree less than 3.  Since these low degree vertices are
prevalent in the WSP graph, we also compare the bias-reduced estimate
to the degree distribution of the WSP graph restricted to vertices of
degree 3 and higher.  Table \ref{table:wspg} shows the unconditioned $\ell_2$ error
for one experiment, and the $\ell_2$ error of the estimated cdfs
conditioned on vertices having degree at least $3$ for a range of
experiments.

\begin{table}[ht]
\begin{center}
\begin{tabular}{|r|r|r|r|r|r|}
\multicolumn{6}{l}{$\Pr[\deg(u) \leq k]$:}\\
\hline
$n$     &$d$
            &$n_m$&$n_t$ & $\text{err}_{\text{biased}}$ 
                                & $\text{err}_{\text{reduced}}$\\
\hline
  4,941 & 2.67 &  2 & $n$ & 0.25 & 0.75 \\
\hline
\multicolumn{6}{l}{$\Pr\left[\deg(u) \leq k\biggiven \deg(u) \geq 3\right]$:}\\
\hline
$n$     &$d$
            &$n_m$&$n_t$ & $\text{err}_{\text{biased}}$ 
                                & $\text{err}_{\text{reduced}}$\\
\hline
  4,941 & 2.67 &  2 & $n/8$ & 0.24 & 0.13 \\
        &      &    & $n/2$ & 0.12 & 0.06 \\
        &      &    & $n$   & 0.06 & 0.05 \\
        &      &  8 & $n/2$ & 0.09 & 0.06 \\
        &      & 16 & $n/2$ & 0.08 & 0.09 \\
\hline
\end{tabular}
\end{center}
\caption{$\ell_2$ error in degree distribution
  estimation  with and without bias reduction for Western States Power
  Graph ($n = 4,941$, $m = 6,594$) with $n_m$ monitors and $n_t$
  targets (median values
  of $100$ trials).}
\label{table:wspg}
\end{table}

\section{AS Graph}
\label{sec:as graph}
The previous two sections showed theoretically and by computer
simulations that the bias-reduction technique developed in Section
\ref{sec:estimation technique} can be an effective way to reduce the
errors introduced by traceroute sampling.  This section reports on the
results of applying the bias-reduction technique to traceroute-sampled
data from the CAIDA skitter project.

A recent paper by Mahadevan, Krioukov, Fomenkov, Dimitropoulos,
claffy, and Vahdat provides a detailed analysis of CAIDA skitter data
from March, 2004 \cite{MahadevanKFDcV2006}.  We follow the methodology
used there, and, in particular, we aggregate the routes observed over
the course of a month (from daily graphs provided by CAIDA), and we
remove all AS-sets, multi-origin ASes, and private ASes, and discard
all indirect links.

The results of applying the bias-reduction technique to the March,
2004 skitter data are plotted in Figure \ref{fig:AS-degree-sequence}.
This data set contains $9,204$ nodes and $28,959$ edges, so the
average degree before bias reduction is $6.29$.  There are 22 ASes in
the monitor set, and between $10\%$ and $50\%$ of ASes are represented in
the target set.  The bias-reduction technique yields an estimate of
$\widehat{\deg}(u) < \infty$ for $2,078$ vertices, and the average
degree after bias reduction is $12.66$ (which is very close to
$12.52$, the biased average degree of vertices with degree at least
$3$).

%There is an interesting discrepancy between degree distribution estimates
%based on traceroute sampling and {\tt WHOIS} lookup.  As reported in
%\cite{MahadevanKFDcV2006}, the degree distribution from the {\tt
%WHOIS} database has a much higher degree than 

The behavior of the bias reduced estimate for $k$ values around $64$
is particularly interesting (see Figure \ref{fig:AS-degree-sequence}).
Although it is far from definitive, the lack of ASes with degree
between 65 and 90 could be the result of economic or technological
factors.  For example, the Juniper T320 edge router has the ability to
house up to 64 interfaces in one chassis.  This, or similar product
specifications, could lead AS operators to avoid connecting to
\emph{slightly} more than 64 other ASes.

\begin{figure*}[t!]
\begin{center}
\begin{tabular}{cc}
\psfrag{True Distribution}[cl][cl]{\tiny True Dist.}
\psfrag{Biased Estimate}[cl][cl]{\tiny Biased Est.}
\psfrag{Biased Reduced}[cl][cl]{\tiny Bias Reduced}
\includegraphics[width=3.2in]{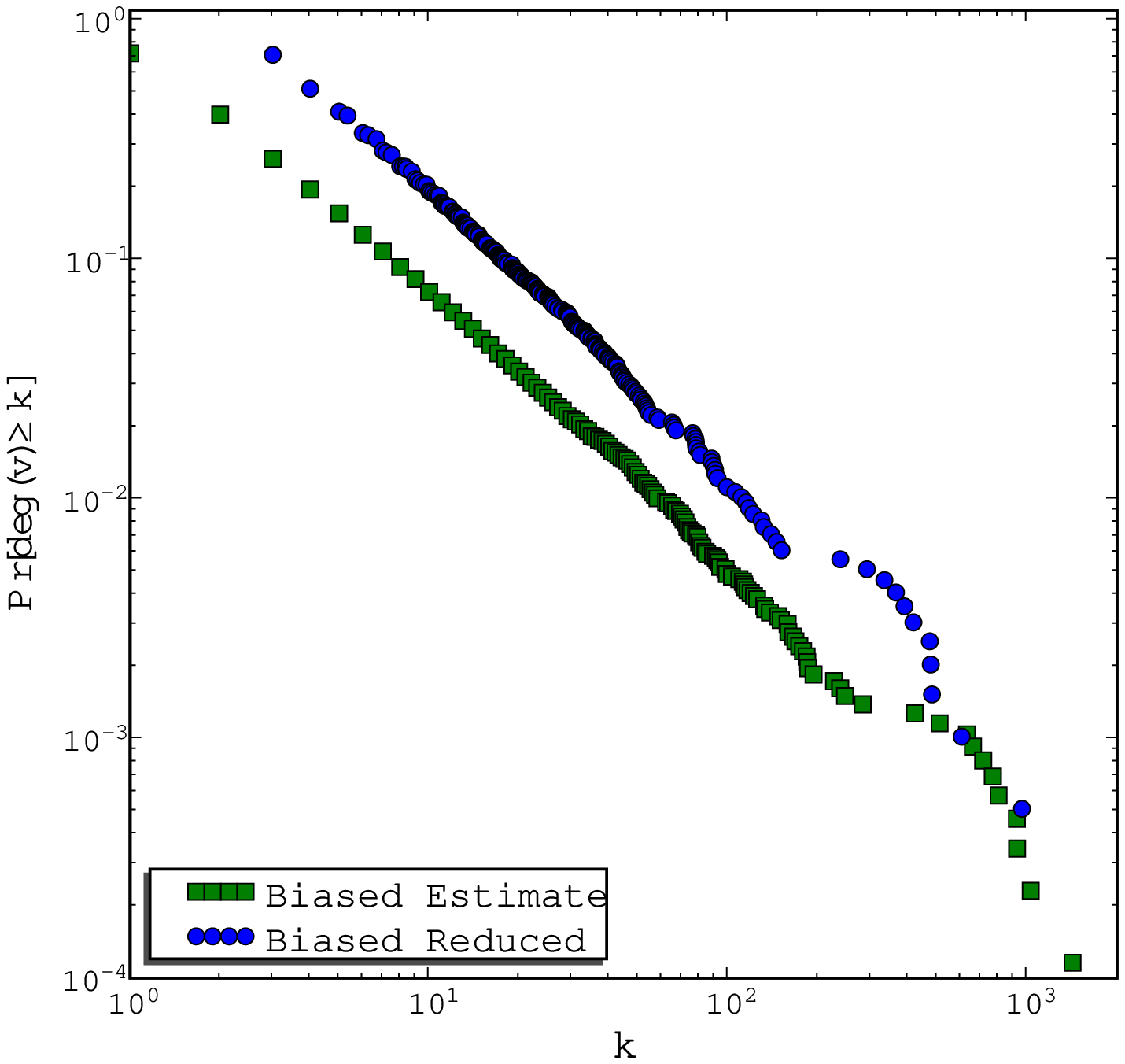}
&
\psfrag{True Distribution}[cl][cl]{\tiny True Dist.}
\psfrag{Biased Estimate}[cl][cl]{\tiny Biased Est.}
\psfrag{Bias Reduced}[cl][cl]{\tiny Bias Reduced}
\includegraphics[width=3.2in]{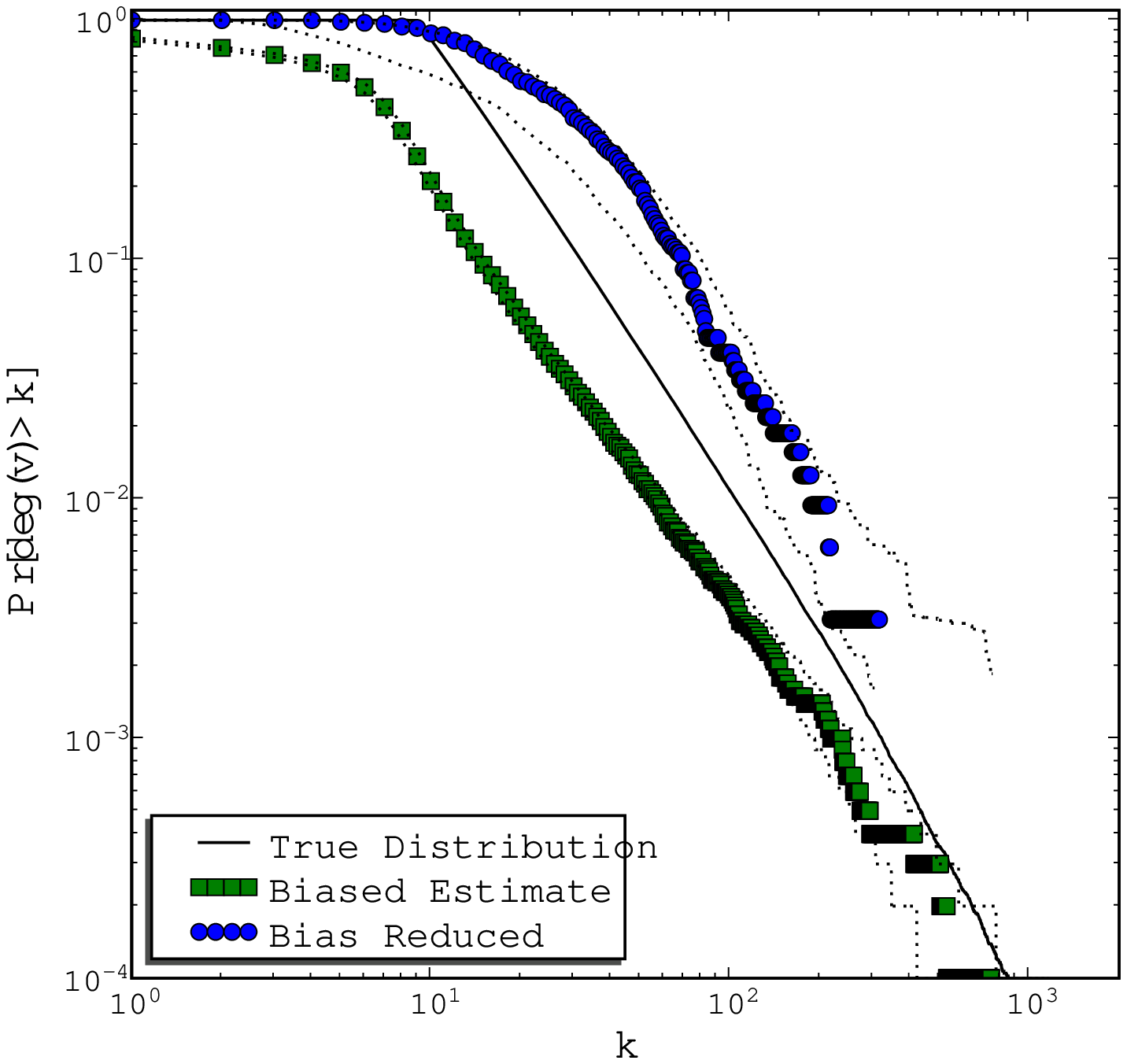}\\
(a) CAIDA skitter, March, 2003.
&
(b) PA graph.
\end{tabular}
\end{center}
\caption{Estimated degree distribution ccdf of CAIDA skitter data from
  March, 2003 with and without bias reduction and estimated degree
  distribution ccdf of PA Graph with similar parameters ($n=10,000$
  nodes, $m=10$ out-edges per node, $20$ source nodes, $n/2$ target
  nodes) with and without bias reduction.  Both estimates of skitter data
  follow power law, but bias reduced estimate of PA Graph does not.}
\label{fig:comp}
\end{figure*}

Finally, the fact that the bias reduced estimate does \emph{not} fall
off at a superlinear rate provides some additional evidence against
the theory that the AS graph is an example of a preferential
attachment model (see comparison in Figure \ref{fig:comp}).  This
argument has been made previously based on completely different
considerations
(see, for example, \cite{ChenCGJSW2002}).

\section{Conclusion}
\label{sec:end}
In this paper we introduced a new approach to addressing the bias inherent
to traceroute sampling.  Starting from the multiple-recapture
population estimation technique of statistics, we developed a bias
reduction technique applicable to the highly dependent random
variables present in path sampling.

In an idealized theoretical framework of shortest path sampling in
\ER graphs, we described how to rigorously prove that the proposed
estimator is asymptotically unbiased, and, using computer experiments,
we show that the estimator can give significant improvements when the
target nodes constitute a fraction of vertex set.  Computer
experiments also highlighted some of the weak points of this estimator,
including the less-than-perfect estimates on locally non-tree-like
graphs, like the PA graph and the random geometric graph.

Applying the bias-reduction technique to the CAIDA skitter data
provided new evidence that the AS graph is not a preferential
attachment graph, and also uncovered a way that economic and
technological limitations are reflected in the AS degree distribution.

The theoretical and computer simulations supporting the effectiveness
of the bias-reduction technique all rely on the assumption that
shortest path routing is a close-enough approximation of BGP routing.
This assumption should be considered in more detail, and the behavior
of the bias-reduction technique under a more realistic model of
traceroute is an important future direction of research.

%Network sampling that is similar to traceroute appears in sociology
%and also in other large computer networks, like the ``Where's George''
%network of dollar bill usage.  Adaption the  bias-reduction technique
%for these setting in another interesting direction for this work.

\section{Acknowledgements}
ADF would like to thank Josh Grubman for pointing us towards the
specifications of the Juniper T320 router, even if he does not believe
that product specifications are likely to result in the absence
of nodes with degree slightly above 64.

\bibliographystyle{abbrv}
\bibliography{../Biblio}

\end{document}